\documentclass[twocolumn,runningheads]{svjour2}
\smartqed  
\usepackage{graphicx}
\usepackage{amssymb}

\journalname{Astrophysics and Space Science}

\def\gtsim{\footnotesize{\mathop{\raisebox{-.4ex}{\rlap{$\sim$}}
\raisebox{.4ex}{$>$}}}}

\begin{document}

\title{Detection potential to point-like neutrino sources with the NEMO-km$^3$ telescope
}


\author{C. Distefano for the NEMO Collaboration}


\institute{C. Distefano \at
              LNS-INFN, via S. Sofia 62, 95123 Catania (Italy) \\
              Tel.: +39 095 542304\\
              Fax: +39 095 542398\\
              \email{distefano\_c@lns.infn.it}
}

\date{Received: date / Accepted: date}

\maketitle

\begin{abstract}
The NEMO Collaboration is conducting an R\&D activity towards the construction of a
Mediterranean km$^3$ neutrino telescope. In this work, we present the results of Monte
Carlo simulation studies on the capability of the proposed NEMO telescope to detect and identify
point-like sources of high energy muon neutrinos.

\end{abstract}

\section{Introduction}
\label{intro}

The detection of high energy neutrinos is considered one of the
most promising means to investigate non-thermal processes in the
Universe. A first generation of small scale detectors has been
realized (NT-200 \cite{Baikal} in the Baikal lake and AMANDA
\cite{Amanda} at the South Pole), demonstrating the possibility to
use the \v{C}erenkov technique to track high energy neutrinos.
Besides, these experiments have set limits on neutrino fluxes.
Other small scale detectors are at different stage of realization
(ANTARES \cite{antares} and NESTOR \cite{nestor}).

Actual expectations on neutrino fluxes, mainly
based on the measured cosmic ray fluxes and the
estimated fluxes for several high energy sources from
theoretical models \cite{flussinu}, require
detectors of km$^3$ scale.
Following the success of AMANDA,
the largest operating detector, the
realization of the IceCube km$^3$ detector \cite{icecube} has started
at the South Pole. On the other hand, many issues, as
the full sky coverage, strongly support the
construction of a km$^3$ scale detector in Mediterranean
Sea.

\section{The NEMO project}

The NEMO Collaboration \cite{NEMO} is performing
R\&D towards the design and construction of the Mediterranean
km$^3$ neutrino detector.
The activity was mainly focused on the search and characterization of an
optimal site for the detector installation and on the
development of a feasibility study for the detector.
A deep sea site with optimal features in terms of
depth and water optical properties has been identified
at a depth of 3500 m about 80 km off-shore Capo
Passero and a long term monitoring of the site has
been carried out \cite{Riccobene05}.
The feasibility study of the km$^3$ detector includes
the analysis of all the construction and installation
issues and the optimization of the detector geometry
by means of numerical simulations. The validation of
the proposed technologies via an advanced R\&D
activity, the prototyping of the proposed technical
solutions and their relative validation in deep sea
environment will be carried out with the two pilot
projects NEMO Phase-1 and Phase-2 \cite{MignecoVLVnT2}.

\section{Detector lay-out}

The geometry of the NEMO-km$^3$ telescope simulated
in this work is a square array of $9\times9$
towers equipped with 5832 optical
modules (10" diameter PMTs) \cite{MignecoVLVnT2}. The towers, moored on the
seabed at 3500 m depth, have an instrumented height of
680 m, with a storey-storey distance of 40 m, and are
spaced of 140 m. The detector response is simulated
using the codes developed by the ANTARES
Collaboration \cite{antares_codes}. In the simulation codes, the light
absorption length, measured in the site of Capo Passero
($L_a\approx 68$ m at 440 nm \cite{Riccobene05}), is taken into account. Once
the sample of PMT hits is generated, spurious PMT hits,
due to the underwater optical noise ($^{40}$K decay), are
introduced, with a rate of 30 kHz for 10"
PMTs, corresponding to the average value measured in Capo
Passero site.

\begin{figure}[h]
\begin{center}
\vspace*{-0.8cm}
\includegraphics[width=7.5cm]{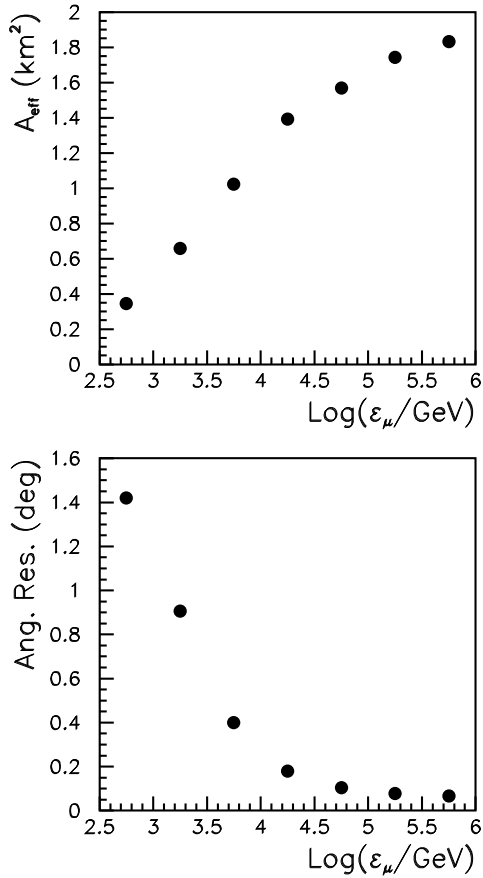}
\end{center}
\vspace*{-0.5cm}
\caption{Effective area and angular resolution of the simulated NEMO detector as
a function of the muon energy. The detector response is simulated considering
a diffuse flux of up-going muons reaching the detector surface.}
\label{fig:aeff}
\end{figure}

The simulated detector lay-out reaches an effective area of 1 km$^2$ at a muon
energy of about 10 TeV and an angular resolution of a few tenths of degrees at the
same energy as plotted in Fig. \ref{fig:aeff}.

\section{Detector pointing accuracy}

The detector angular resolution is one of the most important
parameters in the identification of point-like sources.
Therefore, an experimental determination is required.
A possible method, already
adopted in cosmic ray detectors, consists in the observation of
the so called {\it Moon shadow} \cite{macro}. Since the Moon
absorbs cosmic rays, we expect a lack of atmospheric muons from
the direction of the Moon disk. The detection of the muon deficit
provides a measurement of the detector angular resolution.

\begin{figure}[h]
\begin{center}
\vspace*{-0.8cm}
\includegraphics[width=9cm]{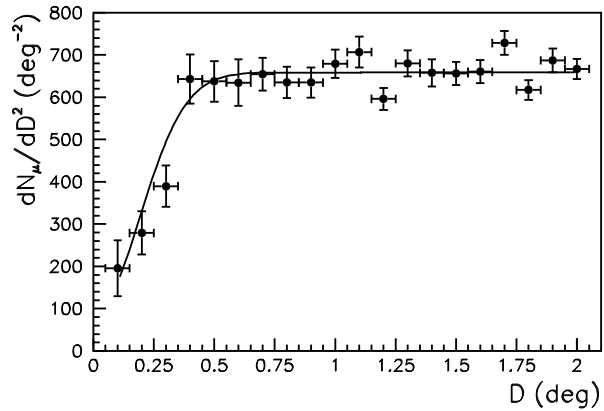}
\end{center}
\vspace*{-0.5cm} \caption{Detected muon event density versus the
angular distance from the Moon center, assuming 1 year of data
taking. The points are fitted assuming a Gaussian shape for the
point spread function, obtaining an angular resolution of
$\sigma=0.19^\circ\pm0.02^\circ$ \cite{VLVnT2}.}
\label{fig:moon_plot}
\end{figure}

Monte Carlo simulations show that the NEMO telescope could be able to
detect the {\it Moon shadow} and that about 100 days are needed to
observe a $3\sigma$ effect. Assuming a detector point spread
function with a Gaussian shape, the same simulations yields a
detector angular resolution $\sigma=0.19^\circ\pm0.02^\circ$
(see Fig. \ref{fig:moon_plot}). Besides detecting its position in the sky allows us
to determine the absolute orientation of the detector
\cite{VLVnT2}.

\section{NEMO-km$^3$ sensitivity to neutrinos from point-like sources}
\label{sec:sensitivity}

\subsection{Calculation of the detector sensitivity}
\label{sec:sensi_def}

The detector sensitivity spectrum is calculated according to the following formula:
\begin{equation}
\left(\frac{d\varphi_\nu}{d\varepsilon_\nu}\right)_{90}=\frac{\overline{\mu}_{90}(b)}{N_\mu}\left(\frac{d\varphi_\nu}{d\varepsilon_\nu}\right)_0,
\end{equation}
where $\overline{\mu}_{90}(b)$ is the 90\% c.l. average upper limit for an
expected background $b$ and calculated as suggested by Feldman and Cousins \cite{Feldman98};
$(d\varphi_\nu/d\varepsilon_\nu)_0$ is an arbitrary point source spectrum
inducing a mean signal $N_\mu$.
The detector sensitivity is calculated taking
into account both atmospheric neutrino and muon backgrounds.

\subsection{Simulation of the atmospheric muon and neutrino background}

A sample of $7\cdot10^9$ atmospheric neutrinos have been generated
using the ANTARES event generation code, based on a weighted
generation technique \cite{antares_codes}. The events were generated in the energy
range $10^2\div10^8$ GeV, with a spectral index $X=2$ and a $4\pi$
isotropic angular distribution. The events were then weighted
to to the sum of the Bartol flux \cite{Agrawal96} and of prompt
neutrino {\tt rqpm} model \cite{Bugaev98} flux. 
When the event weight is calculated, the neutrino
absorption in the Earth, as a function of neutrino energy and direction,
is taken into account \cite{nemo_mqso}.
So doing, we
compute a number $\approx 4\cdot 10^4$ of detected atmospheric
neutrino events per year of data acquisition.

Atmospheric muons are generated at the detector, applying a
weighted generation technique.
We generated a sample of $N_{total}= 2.5\cdot 10^7$ muons, in
the energy range 1 TeV $\div$ 1 PeV, with a generation
spectral index $X=3$. We also generated
$N_{total}= 4\cdot 10^7$ events in the range 100 GeV - 1 TeV,
with a generation spectral index $X=1$.
Muons are generated with an isotropic angular distribution.
The events are weighted to the
Okada parameterization \cite{okada}, taking into account the depth of the
Capo Passero site ($D=3500$ m) and the flux variation inside
the detector sensitive height ($h\approx900$ m).
According to the Okada parameterization, the expected number
of reconstructed muon events is about $4\cdot 10^8$ per year.
Atmospheric muon parameterization by Bugaev et al \cite{igor} is also
considered. In this case the expected number of reconstructed events
is $5\cdot 10^8$ per year but no significant differences are observed in the
detector sensitivity values.

The simulated statistics cover only a few days. Considering that reconstructed events
have a flat distribution in Right Ascension (RA), we can project the
simulated events in a few degrees bin $\Delta$RA, centered in the source position.
So doing, we get statistics of atmospheric muons corresponding to a time
$\gtsim1$ year at all source declinations.

Major details on the Monte Carlo simulation of the NEMO telescope could
be found in ref. \cite{nemo_mqso}.

\subsection{Criteria for the atmospheric background rejection}
\label{sec:selection}

The used reconstruction algorithm is a robust track fitting procedure based
on a maximization likelihood me\-thod \cite{antares_codes}.
In this work, we used, as a {\it goodness of fit criterion}, the variable:
\begin{equation}
\Lambda\equiv -\frac{\hbox{log}(\mathcal{L})}{N_{DOF}}+0.1(N_{comp}-1),
\end{equation}
where $\hbox{log}(\mathcal{L})/N_{DOF}$ is the log-likelihood
per degree of freedom ($N_{DOF}$) and $N_{comp}$ is the total number of compatible
solutions found by the reconstruction program. In particular, events are selected if
the variable $\Lambda$ is greater than a given value $\Lambda_{cut}$.
This quality cut is here applied together with other selection criteria as listed in the following:

\begin{itemize}

\item the number of hits $N_{fit}$, used to reconstruct the muon track,
must be greater than a given value $N_{fit}^{cut}$;

\item the muon must be reconstructed with $\vartheta_\mu^{rec}<\vartheta_\mu^{max}$,
in order to reject down-going events;

\item only events reconstructed in a circular sky region
centered in the source position and having a radius of
$r_{bin}$ are considered.

\end{itemize}

The optimal values of $\Lambda_{cut}$, $N_{fit}^{cut}$, $\vartheta_\mu^{max}$ and $r_{bin}$ are chosen to optimize
the detector sensitivity.

\subsection{Detector sensitivity to neutrino point-like sources}

In this section, we calculate the expected detector sensitivity to
neutrinos from point-like sources. We simulated muons induced by
$\sim10^9$ neutrinos with energy range $10^2\div10^8$ GeV and X=1.
These events are weighted to the neutrino spectrum
$(d\varphi_\nu/d\varepsilon_\nu)_0=10^{-7}\varepsilon_{\nu,GeV}^{-\alpha}$
(GeV$^{-1}$ cm$^{-2}$ s$^{-1}$). As a first case, we consider a
source at a declination of $\delta=-60^\circ$. Such a source has a
24 hours of diurnal visibility and it covers a large up-going
angular range ($\vartheta_\mu=24^\circ-84^\circ$). The sensitivity
for this source is therefore representative of the average
response of the NEMO detector.

\begin{figure}[h]
\begin{center}
\vspace*{-0.8cm}
\includegraphics[width=9.5cm]{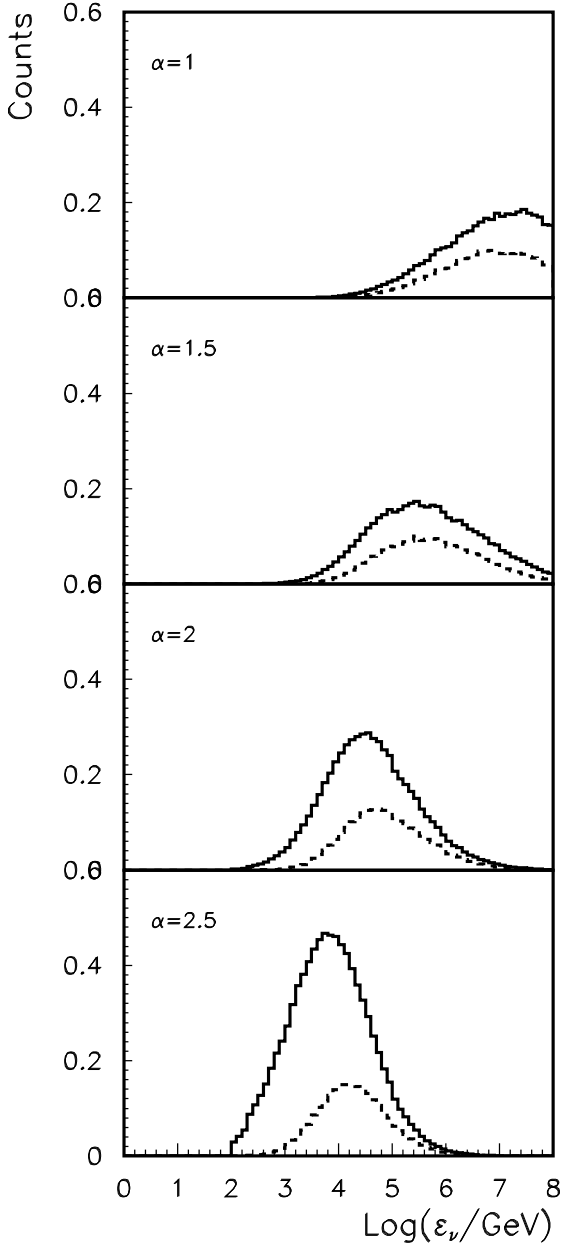}
\end{center}
\vspace*{-0.5cm}
\caption{Neutrino energy spectra of reconstructed
events, before (solid line) and after (dashed line) the event
selection. The spectra are normalized to the sensitivity fluxes for 3 years
of data taking.}
\label{fig:spettri}
\end{figure}

In Table \ref{tab:sens-point-coin-0-180-9x9}, we report the
expected sensitivity for different values of the spectral index
$\alpha$, considering 3 years of data taking. In Fig.
\ref{fig:spettri}, we show how the energy spectrum of
reconstructed events is reduced after the event selection, varying
the neutrino spectral index $\alpha$. The same plots show also as
the spectrum peak moves towards lower energies for softer spectral
indices.

\begin{table}[h]
\caption{Sensitivity to a point-like neutrino source at $\delta=-60^\circ$, for different
spectral indices $\alpha$ and 3 years of
data taking. The sensitivity spectrum $\varepsilon_\nu^\alpha(d\varphi_\nu/d\varepsilon_\nu)_{90}$
is expressed in GeV$^{\alpha-1}/$ cm$^2$ s.}
\centering
\label{tab:sens-point-coin-0-180-9x9}
\begin{tabular}{cccccccc}
\hline\noalign{\smallskip}
$\alpha$ & $\Lambda_{cut}$ & $N_{fit}^{cut}$ & $\vartheta_\mu^{max}$ & $r_{bin}$  &  $\bar{\mu}_{90}(b)$ & $\varepsilon_\nu^\alpha(d\varphi_\nu/d\varepsilon_\nu)_{90}$ \\[3pt]
\tableheadseprule\noalign{\smallskip}
1.0   &   -7.6 &  30 &  90$^\circ$  &   0.4$^\circ$   &           2.4  &     $1.9\cdot10^{-15}$    \\
1.5   &   -7.6 &  30 &  90$^\circ$  &   0.4$^\circ$   &           2.5  &     $2.6\cdot10^{-12}$    \\
2.0   &   -7.3 &  -  &  90$^\circ$  &   0.5$^\circ$   &           2.8  &     $1.2\cdot10^{-9}$     \\
2.5   &   -7.3 &  -  &  90$^\circ$  &   0.6$^\circ$   &           2.9  &     $2.3\cdot10^{-7}$     \\
\noalign{\smallskip}\hline
\end{tabular}
\end{table}

\begin{figure}[h]
\begin{center}
\vspace*{-0.8cm}
\includegraphics[width=9cm]{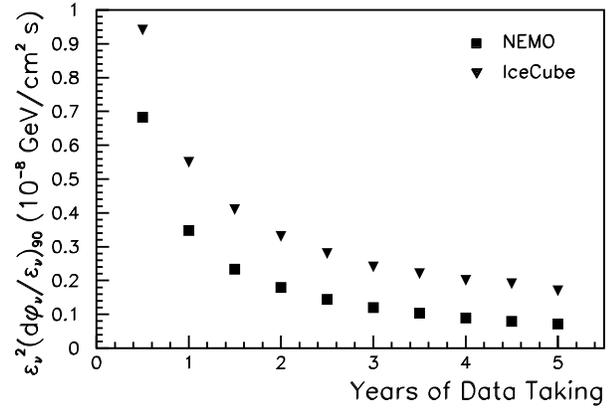}
\end{center}
\vspace*{-0.5cm}
\caption{Sensitivity to a neutrino spectrum with $\alpha=2$,
coming from a $\delta=-60^\circ$ declination point-like
source and comparison with the IceCube detector \cite{icecube}.}
\label{fig:point-sensi}
\end{figure}

The detector sensitivity
was calculated as a function of the years of data taking and reported in Table \ref{tab:nemo20dh140-9x9-years}.
Results for $\alpha=2$ are also plotted in Fig. \ref{fig:point-sensi}, compared to the IceCube sensitivity
obtained for a $1^\circ$ search bin \cite{icecube}.
Our results show that the proposed NEMO detector reaches a better
sensitivity to muon neutrino fluxes with a smaller search bin.

\begin{table}[h]
\caption{Sensitivity to a point-like neutrino source at $\delta=-60^\circ$, for different
spectral indices $\alpha$ and for different number of years of data taking.
The sensitivity spectrum $\varepsilon_\nu^\alpha(d\varphi_\nu/d\varepsilon_\nu)_{90}$
is expressed in GeV$^{\alpha-1}/$ cm$^2$ s.}
\centering
\label{tab:nemo20dh140-9x9-years}
\begin{tabular}{ccccc}
\hline\noalign{\smallskip}
$\alpha$ &  1            &  1.5          &  2           &   2.5         \\[3pt]
\tableheadseprule\noalign{\smallskip}
years    &               &               &              &               \\
 ~1  &  5.9$\cdot10^{-15}$  &  7.8$\cdot10^{-12}$  &  3.5$\cdot10^{ -9}$  &  6.5$\cdot10^{ -7}$  \\
 ~2  &  2.9$\cdot10^{-15}$  &  3.9$\cdot10^{-12}$  &  1.8$\cdot10^{ -9}$  &  3.4$\cdot10^{ -7}$  \\
 ~3  &  1.9$\cdot10^{-15}$  &  2.6$\cdot10^{-12}$  &  1.2$\cdot10^{ -9}$  &  2.3$\cdot10^{ -7}$  \\
 ~4  &  1.5$\cdot10^{-15}$  &  2.0$\cdot10^{-12}$  &  8.9$\cdot10^{-10}$  &  1.7$\cdot10^{ -7}$  \\
 ~5  &  1.2$\cdot10^{-15}$  &  1.6$\cdot10^{-12}$  &  7.2$\cdot10^{-10}$  &  1.4$\cdot10^{ -7}$  \\
 10  &  5.9$\cdot10^{-16}$  &  8.1$\cdot10^{-13}$  &  3.7$\cdot10^{-10}$  &  7.1$\cdot10^{ -8}$  \\
\noalign{\smallskip}\hline
\end{tabular}
\end{table}

\begin{figure}[h]
\centering
\includegraphics[width=9cm]{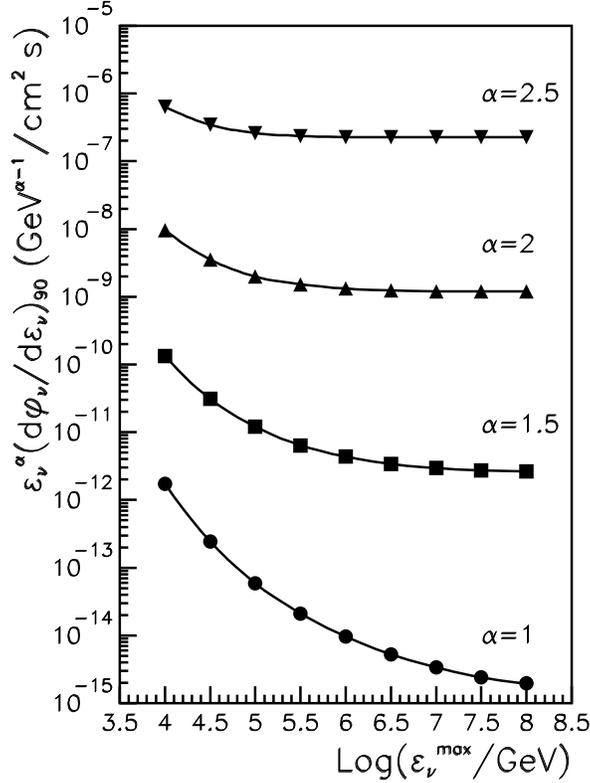}
\caption{Sensitivity to a point-like neutrino source at $\delta=-60^\circ$, for different
spectral indices $\alpha$ and for different values of the high energy cut-off.}
\label{fig:sensi-dec-60-tutte-vere-emax}
\end{figure}

The expected astrophysical neutrino spectra could not extend up to
$10^8$ GeV, especially in the case of Galactic sources. For this
reason, we also computed the detector sensitivity as a function of
the high energy neutrino cut-off $\varepsilon_\nu^{max}$. Results
of our calculations are plotted in Fig.
\ref{fig:sensi-dec-60-tutte-vere-emax}.
Decreasing the energy cut-off, the sensitivity doesn't get worse
until that  $\varepsilon_\nu^{max}$ reaches the energy peak of
reconstructed neutrino spectra (see Fig. \ref{fig:spettri}).
In the case of hard
spectrum sources, the detector sensitivity is better and it gets
better if the spectrum extends to VHE. Expected astrophysical
neutrino spectra could be softer ($\alpha\gtsim2$); in this case
the sensitivity value doesn't vary much with
$\varepsilon_\nu^{max}$.

\begin{figure}[h]
\begin{center}
\vspace*{-0.8cm}
\includegraphics[width=9cm]{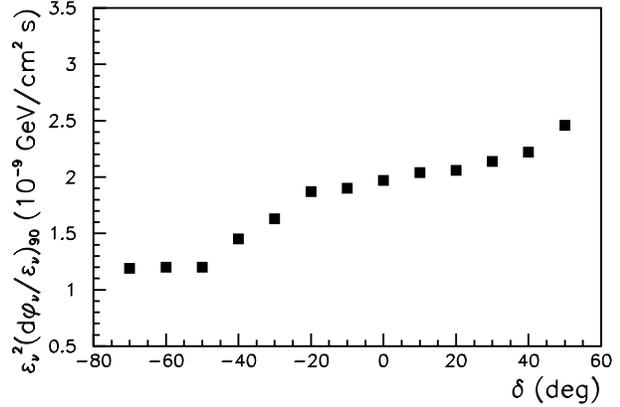}
\end{center}
\vspace*{-0.5cm}
\caption{Sensitivity to a point-like neutrino source at declination $\delta$, for
spectral index $\alpha=2$ and 3 years of
data taking.}
\label{fig:sensi-e2-delta}
\end{figure}

We finally consider the dependence of the sensitivity on the source declination.
In particular, the detector sensitivity gets worse with increasing declination due to the decrease of the
diurnal visibility (to respect with the latitude of the Capo Passero site).
Fig. \ref{fig:sensi-e2-delta} shows the sensitivity versus the source declination, considering
three years of data taking and $\alpha=2$. The worst sensitivity
is $2.5\cdot 10^{-9}$ GeV/cm$^2$ s, calculated for a source declination of $\delta=50^\circ$
for which the diurnal visibility reduces to a few hours per day.

\section{Physics cases}

In this section, we consider the case of two particular sources:
microquasar LS 5039 and SNR RX J1713.7-3946, both observed in the
TeV gamma-ray region. For each source, we compute the detector
sensitivity and the expected number of source events compared with
the background.

\subsection{Microquasar LS 5039}

The H.E.S.S. telescope has recently detected TeV $\gamma$-rays from LS
5039  \cite{hess-ls5039}. This discovery provided the first unambiguous
evidence for presence of multi-TeV particles in microquasars.

Aharonian et al. \cite{Aharonian05} discussed different possible
scenarios for the production of the observed $\gamma$-ray flux.
They considered both leptonic and hadronic production mechanisms
and argued in favor of a TeV photon flux originating from pp
interaction. If so, $\gamma$-rays should be accompanied by TeV
neutrinos with an average energy flux of $f_\nu^{th}=10^{-10}$ erg/cm$^{2}$
s.

In Tab. \ref{tab:ls5039} are summarized the detector sensitivities
for microquasar LS 5039, assuming neutrino fluxes with spectral
indices $\alpha=1.5$ and 2, in the energy range 0.1 TeV and
$\varepsilon_\nu^{max}=10$ and 100 TeV. The expected number of
selected events, induced by the flux $f_\nu^{th}$, is
given in the same table. The comparison with
the atmospheric background shows that an evidence could be
expected in a few years of data taking.

\begin{table}[h]
\caption{Detector sensitivity to neutrinos from LS 5039. The sensitivity
$ f_{\nu,90}$ is expressed in units of erg/cm$^{2}$ s and refers
to a detector live time of 3 years.
The corresponding values of $\vartheta_\mu^{max}$, $\Lambda_{cut}$ and $r_{bin}$
are also given. We also report the number N$_\mu^{m}$ from the source compared to the atmospheric background events N$_\mu^{b}$
surviving the event selection.}
\centering
\label{tab:ls5039}
\begin{tabular}{cccccc|cc}
\hline\noalign{\smallskip}
$\alpha$ & $\varepsilon_{\nu,TeV}^{max}$  &  $\Lambda_{cut}$ &    $\vartheta_\mu^{max}$ & $r_{bin}$ &   $ f_{\nu,90}$   & $N_\mu^{s}$ &  $N_\mu^{b}$\\[3pt]
\tableheadseprule\noalign{\smallskip}
  1.5            & ~10  & -7.3   &   ~99$^\circ$       & 0.6$^\circ$  & 6.5$\cdot10^{-11}$  & 4.3      & 0.4      \\
  1.5            & 100  & -7.2   &   102$^\circ$       & 0.5$^\circ$  & 2.0$\cdot10^{-11}$  & 13.1     & 0.2      \\
  2.0            & ~10  & -7.3   &   ~99$^\circ$       & 0.7$^\circ$  & 1.2$\cdot10^{-10}$  & 2.5      & 0.6      \\
  2.0            & 100  & -7.3   &   102$^\circ$       & 0.5$^\circ$  & 3.8$\cdot10^{-11}$  & 7.1      & 0.3      \\
\noalign{\smallskip}\hline
\end{tabular}
\end{table}

\subsection{SNR RX J1713.7-3946}

The CANGAROO Collaboration observed $\gamma$-rays from SNR RX
J1713.7-3946, claiming the hadronic origin of the measured energy
spectrum \cite{Enomoto02}. Alvarez-Mu\~{n}iz and
Halzen (A\&H) \cite{Alvarez02} calculated the high-energy neutrino flux
associated with this source. Their calculations yield to an
expected neutrino spectrum
\begin{equation}
\varepsilon_\nu^\alpha(d\varphi_\nu/d\varepsilon_\nu)^{th}=4.14\cdot10^{-8}
\hbox{ cm$^{-2}$ s$^{-1}$ GeV$^{-1}$},
\end{equation}
with spectral index $\alpha=2$ and extending up to $\sim$ 10 TeV.
More recent calculations were performed by Costantini and Vissani (C\&V)
\cite{Costantini05}, based on the $\gamma$-rays flux measured by
the H.E.S.S. experiment \cite{hess-snr}. According to their
calculations, we expect a neutrino spectrum
\begin{equation}
\varepsilon_\nu^\alpha(d\varphi_\nu/d\varepsilon_\nu)^{th}=3\cdot10^{-8}
\hbox{ cm$^{-2}$ s$^{-1}$ GeV$^{-1}$},
\end{equation}
with $\alpha=2.2$ and a neutrino energy ranging between 50 GeV and
1 PeV.

In this paper, we calculate the expect detector sensitivity for RX
J1713.7-3946. Despite it is an extended source, in a first
approximation we can consider it as point-like since its diameter
($\varnothing=1.3^\circ$) \cite{hess-snr} is comparable o smaller
than the detector search bin. Results are reported in Tab.
\ref{tab:rxj1713}, where the dependence of the sensitivity on the
spectral index $\alpha$ and on the high energy cut-off
$\varepsilon_\nu^{max}$ is shown. In the same table, we also
report the number of expected events surviving the selection
criteria, considering both the theoretical predictions. The
comparison with the atmospheric background shows that, also in
this case, the NEMO telescope could identify the source in a few years
of data taking.

\begin{table}[h]
\caption{Detector sensitivity to neutrinos from the SNR RX J1713.7-3946.
The sensitivity is expressed in units of cm$^{-2}$ s$^{-1}$ GeV$^{-1}$ and refers
to a detector live time of 3 years.
The corresponding values of $\vartheta_\mu^{max}$, $\Lambda_{cut}$ and $r_{bin}$
are also given. We also report the number N$_\mu^{m}$ from the source compared to the atmospheric background events N$_\mu^{b}$
surviving the event selection.}
\centering
\label{tab:rxj1713}
\begin{tabular}{ccccc|cc}
\hline\noalign{\smallskip}
Model  &  $\Lambda_{cut}$  &   $\vartheta_\mu^{max}$ & $r_{bin}$ &  $\varepsilon_\nu^\alpha(d\varphi_\nu/d\varepsilon_\nu)_{90}$  &
$N_\mu^{s}$ &  $N_\mu^{b}$\\[3pt]
\tableheadseprule\noalign{\smallskip}
  A\&H      & -7.3  &  ~99$^\circ$ & 0.6$^\circ$  &  1.4$\cdot10^{-8}$  & 8.5 &      0.6   \\
  C\&V      & -7.3  &  101$^\circ$ & 0.4$^\circ$  &  1.7$\cdot10^{-8}$  & 4.8 &      0.4    \\
\noalign{\smallskip}\hline
\end{tabular}
\end{table}

\section{Conclusions}

The possibility to detect TeV muon neutrinos
from point-like sources with the proposed NEMO-km$^3$ underwater
\v{C}eren\-kov neutrino telescope has been investigated.
In particular Monte Carlo simulations were carried out to determine
the expected response of the km$^3$ telescope.

Our simulations show that it could be possible to observe the Moon
shadow in about 100 days. This detection provides a measurement of
the detector angular resolution and of the detector pointing
accuracy.

We also computed the detector sensitivity to muon neutrinos from point-like
sources, defined as the minimum flux detectable with respect to the
atmospheric muon and neutrino background.
The dependence of the sensitivity on the neutrino spectral index
and energy range, on the source declination and on years of data taking
has been studied.

Finally, we consider the case of two particular sources:
microquasar LS 5039 and SNR RX J1713.7-3946; both observed in the
TeV gamma-ray region. For each source, we compute the detector
sensitivity and the expected number of source events compared with
the background.
Our results show that,
assuming present predictions of TeV neutrino fluxes,
the proposed NEMO telescope could identify both sources in a few years of data taking.



\begin{thebibliography}{3}
\bibitem{NEMO}
The NEMO web site: nemoweb.lns.infn.it.
\end{thebibliography}
\end{document}